# A Technique for the Detection of PDF Tampering or Forgery

Gabriel Grobler, Sheunesu Makura[2][0000-0002-5129-3216] and Hein Venter[3][0000-0002-3607-8630]

University of Pretoria, Hatfield 0028, South Africa
[1]u20534541@tuks.co.za, [2]makura.sm@up.ac.za, [3]hein.venter@up.ac.za

**Abstract.** Tampering or forgery of digital documents has become widespread, most commonly through altering images without any malicious intent such as enhancing the overall appearance of the image. However, there are occasions when tampering of digital documents can have negative consequences, such as financial fraud and reputational damage. Tampering can occur through altering a digital document's text or editing an image's pixels. Many techniques have been developed to detect whether changes have been made to a document. Most of these techniques rely on generating hashes or watermarking the document. These techniques however have limitations in that they cannot detect alterations to portable document format (PDF) signatures or other non-visual aspects, such as metadata. This paper presents a new technique that can be used to detect tampering within a PDF document by utilizing the PDF document's file page objects. The technique employs a prototype that can detect changes to a PDF document, such as changes made to the text, images, or metadata of the said file.

**Keywords:** PDF, Tampering, Forgery, Metadata, Hashing, Alterations, File page objects

## 1 Introduction

Digital media is a common form of communication and entertainment in modern society. Given the widespread use of these digital forms of media through images, videos, and text-based documents, it has opened a new avenue for tampering with official or unofficial content [1]. Forgery is when a person creates a copy of the original, often intending to commit fraud. Tampering is when a person interferes with or changes the original without permission. In the context of this paper, both terminologies will be used interchangeably.

The tampering that the average person will generally encounter is focused on the alteration of images for the purpose of looking better in posted photos or videos through techniques such as deep fakes [2], where a deep fake is a fake image of an event produced using deep learning techniques. Though many alterations are not of a criminal nature there are alterations of more formal digital documents that could have adverse consequences.

One of the most common forms of digital documents used for communications is the portable document format (PDF). A PDF is a file format developed by Adobe that preserves the layout and formatting of a document regardless of the software, hardware, or operating system used to view or print it. PDF files can contain text, images, forms, annotations, and other data, and they are widely used for distributing documents that need to be displayed and printed consistently across different platforms. Given that the PDF is difficult to alter, it is used in many formal forms of communication, such as memorandums, contracts, specifications, etc. However, with the advancements in technology and the expanded ease of access to tools such as Adobe Acrobat or one of the many free online editors that make altering documents in this format possible, it has become simpler to alter PDF files with limited knowledge of the PDF format. This leads to the necessity to detect and analyze any alterations made to a document in PDF format.

Most of the current techniques that are used for the detection and partial analysis of any alterations use a form of watermarking. Though we will highlight two forms of watermarking in this paper, there are numerous ways of implementing such a watermark into a PDF document or most other forms of digital content.

In addition to watermarking, another technique used to check whether the current PDF matches the original is hashing. Though these techniques successfully detect changes to a PDF document and its contents, they mostly rely on the visible aspects of the said PDF, such as the images and text that a reader would see. This is because they mainly rely on using the visible content to the reader. The techniques discussed would not detect changes to this PDF document's metadata or background data. Should alterations be made that would embed malware into the PDF using the scripting abilities of the PDF format, the above techniques would not be able to detect such changes to the document [7]. Nor would they be able to detect changes to a PDF signature, which could have severe consequences should a PDF document be attacked in such a manner [8]. PDF signatures are elaborated in section 2.3.

It is worth noting that while the existing methods of using watermarking and hashing have their limitations, changes can be identified even though the exact point of change may not be identifiable. This is because process of changing anything in a document creates a different hash for the document which therefore makes it difficult to pinpoint exactly where and what caused the change.

### 1.1   Problem Statement and Research Questions

The problem that this paper is addressing is the tampering or forgery of PDF documents. With the PDF format being used as a formal means of communication in multiple industries, it has become a good target for criminals who wish to affect contracts or aid in misinformation. The adverse effects of such documentation being tampered with or forged could greatly impact many people's lives if it goes undetected. This paper presents a new technique that can detect tampering or forgery of a PDF document using the underlying content of a PDF document, such as the file page objects. File page objects specifically pertain to the content and structure of individual pages within the PDF document.

The main research problem described above can be expanded by asking the following research questions (RQs). These questions are used to assess whether the proposed solution successfully detects tampering or forgery of a PDF document.

**RQ.1 What are the current techniques that are used or have been researched for the detection or tampering of PDF documents?**
This question aims to address what are the current techniques that are employed for the detection of tampering or forgery in PDF documents. In addition, we would want to find out the current state regarding the detection of tampering or forgery within a PDF document.

**RQ.2    Will using the file page objects to generate a hash detect tampering or forgery of a PDF document's text?**
This question aims to address whether using the file page objects of a PDF document to generate a hash value will be able to detect tampering or forgery of the document. It applies in two parts: the first using the rewrite approach or the second using the incremental approach. The technique should be successful regardless of how the changes are made to the PDF document.

**RQ.3    Can a prototype be developed to detect alterations made to an image within the PDF document?**
This question addresses whether the proposed prototype, which is the main contribution of this paper detects alterations made to an image in a PDF document. If the image is altered by methods such as Photoshop or completely replaced, the prototype should be able to detect that a change has been made to such images in the PDF document.

The remainder of the paper is structured as follows: section 2 highlights literature survey of the current techniques used in detecting forgery or tampering, section 3 elaborates on the prosed prototype design and implementation, section 4 discusses the experiments conducted, the results and evaluation then section 5 concludes the paper.

## 2    Literature Review

This section provides a literature survey of the state of the current research conducted on the topic of detecting tampering or forgery within PDF documents.

## 2.1    Watermarking

A watermark is a technique used to protect an original piece of work from being copied. When applied to PDF documents, they are not as visible as a watermark on an image; they are generally embedded and hidden from view. Research by Khadam et al. [5] used a watermark on PDF documents using file page objects from a PDF document. The technique was designed to be non-intrusive, whereby the watermark was embedded into the PDF document to be used for later comparison. The watermark was designed to be fragile, meaning it becomes invalid if an attempt is made to alter the document. While this watermarking technique presented by the authors successfully detects changes to a PDF document's format without bloating the file's size, it does not discuss the idea of detecting whether JavaScript has been embedded into the document.

Research by Usop [4] presented a watermarking technique where the PDF file was first converted to an image in the Bitmap (BMP) format, initiating the watermark creation process. After dividing the image pixels into blocks, these blocks were used in a zigzag manner to create watermarks per block. This method of watermarking efficiently detects changes to the visual appearance of the document. However, it cannot detect alterations that do not affect the document's visual appearance to the reader, such as embedding JavaScript code or tampering with the PDF file signature. Another issue is that generating these watermarks can take a long time, but the high detection accuracy justifies the duration.

Dikanev et al. [9] generated a semi-fragile watermark to protect and detect any alterations made to a PDF document. The authors used Quick Response (QR) codes to generate the watermarks. Initially, the page is converted into a raster image, a simple pixel map, which is compressed before being used to generate the QR code. This QR code is then normalized into a specified range before being embedded into the image of the PDF page using an inverse function. When it is time to verify the QR code of the PDF document, the inverse of the process described above is performed. This technique makes it easy to detect changes to the visual appearance of a PDF document and makes it simple to localize where these changes were made. However, should an individual decide to alter the metadata or embed JavaScript code into a PDF document, this technique would be unable to detect such actions.

Research by Jiang et al. [13] presented a technique for embedding encrypted watermarks into various objects within the PDF file, such as text, images, and forms, ensuring resilience against multiple types of attacks like text editing, format modification, and page extraction. The authors used a tamper detection algorithm to verify the integrity of the content and identify any tampered areas. Their technique was tested on various PDF files, demonstrating high performance in terms of imperceptibility, capacity, robustness, and compatibility compared to existing methods [15].

### 2.2 Hashing

Hashing is an integrity validation technique that relies on one-way hash functions. These functions take an input and produce a string value of what appears to be random characters, which are calculated based on the input.

A hash algorithm was used by Senkyire and Kester [6] as the primary component of their method to detect tampering or forgery of a document. The authors used the SHA-384 algorithm to calculate hashes for the specific input. The SHA-384 algorithm utilizes blocks of size 1024 bits for generating its hashes, which are padded if the data is not the correct size. The process for generating the hash is as follows: the data is divided into n blocks, and the first block is hashed and fed into the next stage of the process. The second block is hashed using the previous block's hash and contents, which are then fed into the following block for hashing. This process repeats until the n-th block. The hash produced by the last block is used for comparison purposes [6]. This method can be applied to a PDF document by converting its pages to images and following the above process.

Our proposed prototype, presented in Section 3, will use hashing to assess PDFs for forgery. The generated hashes can be checked to determine if any changes were made to the PDF.

### 2.3 PDF Signatures

PDF signatures (or digital signatures) are methods used to validate the origin of a PDF document [8]. They verify the authenticity and integrity of the PDF. For a digital signature to be generated for a PDF document, incremental saving must be used so that the PDF signature can be appended as an object to the document, much like an addition or change to the original document. A single document can have multiple signature objects, allowing it to be signed by multiple stakeholders and verifying that those signatures are valid.

The main concept of the paper by Mladenov et al. [8] revolved around exploiting a PDF document without invalidating the PDF signature. Consequently, changes can be made to the document without the person being able to repudiate it. How this is achieved varies but can broadly fall under the following categories: Universal Signature Forgery, Incremental Saving Attack, or Signature Wrapping Attack [8]. These categories broadly aim to make the PDF signature considered valid. Methods used to protect against such attacks focus solely on the PDF signature and are therefore difficult to expand to apply to the rest of the PDF document. Using a signature may validate the origin of the document, but the technique of using file page objects to generate a hash, allowing detection of tampering and semi-localizing where it happened, will aid in validating the document's integrity.

The concept of shadow attacks on PDF signatures was presented by Mainka et al. [10]. The authors looked at PDF attacks that comply with the PDF standard. While most attacks on PDF signatures rely on creating malformed incremental updates, shadow attacks use well-formed incremental updates. Three forms of shadow attacks are discussed: Hide, Replace, and Hide-and-Replace [10]. A Hide attack attempts to conceal relevant content behind a visible layer, such as text behind an image. The second attack, Replace, uses an incremental update to overwrite the previously declared object. Finally, a Hide-and-Replace attack relies on sending the document to contain hidden descriptions of another document. Once they receive the signed document, they append an Xref table that references this other document.

### 2.4 Hiding Malicious Content

The premise behind hiding malicious content relies on using a dual file [7]. A dual file is a file that combines two different file formats into a single entity. Popescu [7] used dual file combines a PDF and Tag Image File Format (TIFF) file. TIFF files are generally used for editing and manipulating high-resolution images. When the victim receives the document, they see the PDF version and sign it, relying on the fact that PDF signatures will not be invalidated when this dual file is converted into the TIFF format. Once an attacker converts the file into this format, they can alter the document's contents, and the document will still have a valid PDF signature. This attack is possible because a TIFF header can be placed within the header of a PDF file without causing issues.

Popescu [7] describes what is known as a Dali attack. This attack leverages the concept of a polymorphic file that combines both PDF and TIFF formats. The attacker hides malicious content within the TIFF part of the file. When the file is initially viewed, only the benign PDF content is visible. However, once the digital signature is applied, the file can be manipulated to reveal the hidden malicious content without invalidating the digital signature.

The concepts of embedding content in places not checked by watermarking techniques and other hashing techniques show that using hashing on the file page objects alone will be insufficient to detect all alterations. This therefore prompts the need for new techniques that can be utilized to other unique objects of a PDF document, such as the signature, the header, and possibly the cross-reference table itself.

### 2.5 Other techniques

Research work by Guangyong et al. [11] presented a novel approach to protecting and tracing the copyright of PDF files using blockchain technology. The authors proposed a blockchain-based copyright protection technique that combines blockchain with data hiding in PDF files. This scheme involves: (i) a new information hiding algorithm that embeds copyright information in PDF files without altering their appearance, (ii) smart contracts for access control and on-chain proofs of PDF file ownership [11]. The experimental evaluation showed that the proposed technique was effective in terms of security and traceability. The data hiding method did not affect the visual quality of the PDF, and the blockchain-based system ensures reliable copyright management.

Research work by Nguinabe et al. [12] introduced a novel approach to detect falsified PDF documents using graph isomorphism. Their technique involved transforming a PDF document into a graph where nodes represent words and edges represent the semantic relationships. The goal was to find an isomorphism between the original and the altered PDF document graphs. Their technique demonstrated 90% accuracy in detecting falsifications, proving its robustness against insertion, deletion, and modification attacks.

## 3 Methodology and Prototype design

This section discusses the prototype's design and the details of its implementation. An initial discussion regarding the PDF structure is provided before delving into the functionality of the prototype components and their interactions.

### 3.1 PDF Structure

A PDF document comprises four main sections [3]: the header, body, cross-reference table, and trailer.

- **Header:** Specifies the version number of the PDF format that the document conforms to.
- **Body:** Composed of references to objects that make up the content of the document.
- **Cross-reference table:** Contains a list of references to objects in the file to allow for random access and reuse of these objects.
- **Trailer:** Provides quick access to the cross-reference table and certain special objects, as PDFs should be read from the end.

There are two ways to update a PDF document. The first is to have the computer rewrite the whole file upon saving and overwriting the old document. The second is to use incremental updates, in which case the additions are added to the end of the file after the trailer, using additional body, cross-reference, and trailer sections for the update.

A PDF document is structured as a collection of objects. These objects are organized into a hierarchy and are interconnected to form the document. File page objects specifically pertain to the content and structure of individual pages within the PDF document. File page objects are important because they encapsulate all the details needed to render and interact with a particular page within the PDF document. They are integral to the rendering process and are essential for any operations involving manipulation, display, or analysis of the document [3] [[5]. File page objects consist of the following components:

- **Content Stream:** The content stream is a sequence of instructions describing how to display the page. It typically includes instructions for drawing text, images, and graphics on the page. We use the content stream in our proposed prototype to generate a Merke Tree (refer to section 3.2).
- **Resources:** These are objects that define the resources available to the page, such as fonts, images, and other reusable assets.
- **Media Box:** Defines the boundaries of the physical medium on which the page is to be printed.
- **Crop Box:** Defines the region to which the contents of the page should be clipped.
- **Rotation:** Specifies the rotation angle of the page.
- **Annotations:** Optional elements that allow for interaction with the user or define actions to be taken when the document is opened, or certain events occur.

### 3.2    Implementation Details

Our proposed prototype can be used in two different ways with a PDF document. Firstly, to protect a PDF document using our developed technique, the PDF must be run through the prototype so that a hash can be generated and inserted into the PDF document. Secondly, to assess a PDF document for forgery or tampering, the PDF must be run through the prototype, and the detection process will determine if changes have been made.

The prototype was implemented using the Python language. The hashlib and Merkly libraries are used to generate the hashes. These two main processes can be run independently of each other. To read the file page objects of the PDF, we make use of the Portable Document Format Read and Write (PDFRW) library. The PDFRW library gives us the ability to read and write PDF files. PDFRW was chosen because it offers lower-level access to PDF structures, which can be beneficial for custom manipulation tasks. Additionally, for certain operations, PDFRW can be faster due to its low-level access, especially when handling large PDFs or complex tasks. Figure 1 outlines the general process flow of the prototype.

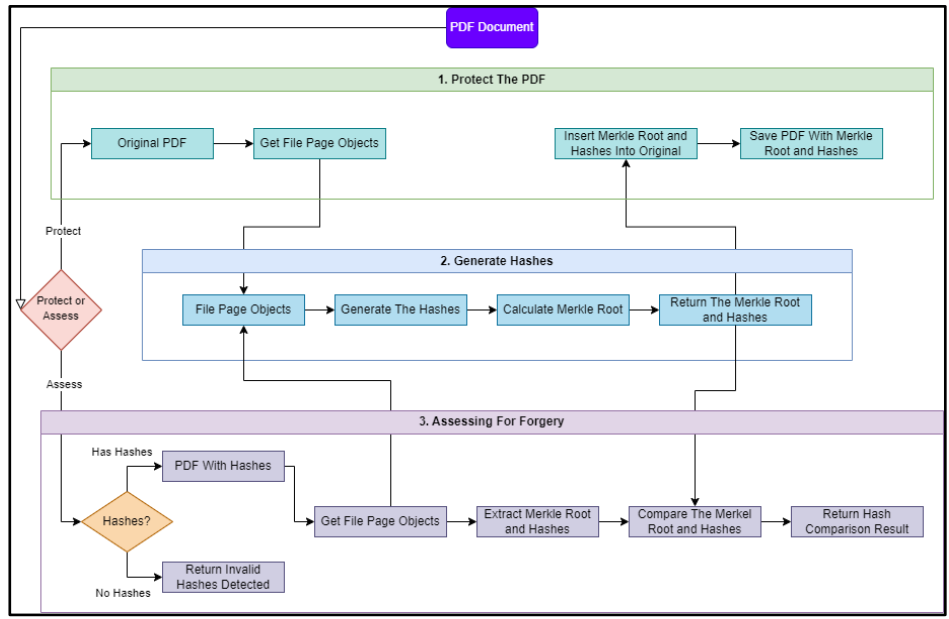

**Fig. 1. Prototype Flow**

The three main sub-components of the prototype will be discussed in more detail in the following sub-sections: in sub-section 3.2.1, we discuss how a PDF document is protected against tampering; in sub-section 3.2.2, we discuss how the hashes are generated for a PDF document; and in sub-section 3.2.3, we discuss how a PDF document is assessed for tampering.

### 3.2.1 Protect the PDF
The PDF document must have the calculated hashes to ensure the technique can detect and analyze tampering or forgery. The first phase, the "Protect The PDF" sub-component from Figure 1, is responsible for this process.

First, it will read the PDF document into the prototype using the PDFRW library, which produces a dictionary-like object for the parts of a PDF document. Once the PDF document has been read, we isolate the file page objects to use them for further processing. An iterative process of computing all the necessary hashes for a particular file page object will begin. For each page, the content stream of the file page object is then used to generate a Merkle tree. This is discussed in detail in section 3.2.3.

Following the calculation of the Merkle tree for the file page object content stream, a hash will be computed using the file page object itself. This is handled using the "2. Generate Hashes" sub-component showed in Figure 1, which is discussed in section 3.2.2. Before the hash can be computed for the file page object, it is first serialized and converted into a format that can be encoded into bytes. Serialization is the process of converting the file page objects and the root object of the PDF into a format that can be easily stored or transmitted, often as a sequence of bytes. Serialization is necessary to ensure that these objects are in a consistent and standard format that can be processed correctly when generating a hash. Some sub-objects of the file page object are excluded during serialization to prevent the calculation of a protected PDF from producing a false result, due to not all PDF document editors creating and updating a PDF document uniformly.

After the hash values are calculated, they are stored in an object structured as follows:
```
{'object': hash value, 'root': hash value, 'leafs': list of hash values}
```
This object is discussed further in detail in section 3.2.2. The values associated with each heading in the object are the values that will be stored within the PDF document. This is done by creating three new keys in the relevant file page object and storing the relevant value inside its respective key. The keys created within the file page object are: 'hashobject', 'hashroot', and 'hashleafs'. If there are four pages within the PDF document, this process will be repeated for all four pages.

The next step to protect the PDF involves generating a hash value for the root object, which is structured as follows:
```
{'root': hash value, 'info': hash value}
```
This object will be discussed further in section 3.2.3. The values associated with each heading in the object are the values that will be stored within the PDF document in the root object. This is done by creating two new keys in the root object and storing the relevant value inside its respective key. The keys created within the file page object are: 'hashroot' and 'hashinfo'.

Once the hashes have all been inserted into the relevant file page objects and the root object, we use the PDFRW library to save a new PDF document. The new PDF document is now the protected PDF document and can be used in the "3. Assessing for Forgery" sub-component. Figure 2 shows an example output of the prototype after the protecting the PDF phase.

```
Would you like to protect a PDF? (y/n): y
Enter the path to the PDF you would like to protect: ./Test_PDFs/Demo.pdf
Protecting: ./Test_PDFs/Demo.pdf
PDF Protected successfully, and saved to ./Test_PDFs/Demo_hash.pdf
Process Completed
```

**Fig. 2. Outcome after successfully protecting a PDF**

### 3.2.2 Generating the Hashes

The second phase, the "2. Generate Hashes" sub-component, is used in sub-components 1 and 3 (refer to Figure 1). It is used to generate the object that was discussed in sub-section 3.2.1. This object is used to protect and assess tampering within the PDF document. The elements of the object and how they are generated is discussed based on the order they appear within the object. For reference, the object looks like as follows:

```
{'object': hash value, 'root': hash value, 'leafs': list of hash values}.
```

The 'object' hash value is calculated using the hashlib library's implementation of the SHA256 hashing algorithm. It is computed using the file page object after it has been serialized and encoded into a byte stream.

The 'root' and 'leafs' hash values are calculated simultaneously using the Merkly library's implementation of a Merkle root. This library utilizes the Keccak hashing algorithm. The Keccak hashing algorithm, also known as SHA-3 (Secure Hash Algorithm 3), is a cryptographic hash function designed to provide high security against various types of attacks, including collision, pre-image, and second pre-image attacks [16]. A Merkle root is the cumulative hash of all the hashes that comprise a Merkle tree. An example is seen in Figure 3. The 'root' hash value that we store within the PDF is the Merkle root of the Merkle tree is created using the content stream within the file page object.

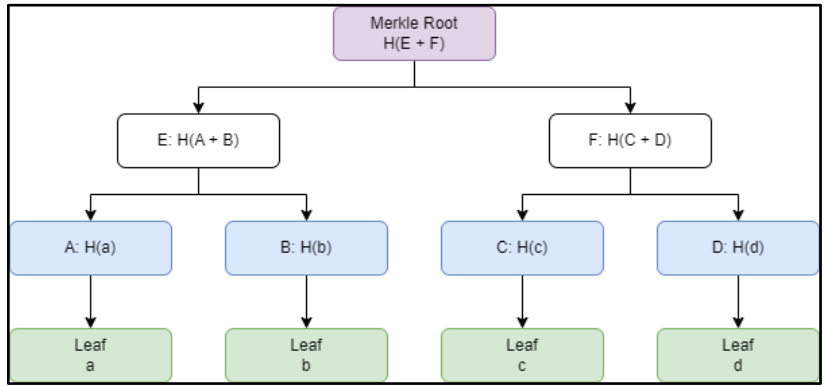

**Fig. 3. Structure of the Merkle Tree**

The content stream is divided into groups of 256 bytes, which are then used as the leaves within the Merkle tree. The 'leafs' hash values in the object are the previously calculated hash values of these groups. These values are stored within the PDF to localize where a change has been made within the file page object. If a change is made to the content of the file page object, the change can be localized to the nearest 256-byte group of the content stream.

Once all three groups of hash values have been calculated, they are returned to the sub-component that called the second phase, the "2. Generate Hashes" sub-component. This could be phase 1 or phase 3.

After the previously mentioned hash values have been stored in the PDF, the calculation of the hash values for the root and metadata will be done. This is done second to the hash values for the file page objects because those hashes are included when calculating the root hash value. The hash object that will be produced is:
{'root': hash value, 'info': hash value}.

The root hash value is calculated by serializing the sub-objects of the Root and calculating the hash with the produced value using the SHA256 algorithm. The info hash value is calculated by serializing the metadata of a PDF document and again using the SHA256 algorithm. The hash object is returned so the values can be stored in the root object or used for comparison. Figure 4 shows the output of the generated hashes.

**Fig. 4. Output of Generated Hashes**

The following sub-section discusses the process of assessing a PDF for tampering or forgery.

### 3.2.3 Assessing for PDF forgery

The second primary function that the prototype can accomplish is the assessment of tampering within a PDF document. For phase 3, "3. Assessing for Forgery", to successfully detect tampering, it requires that phase 1 be run on the PDF document before the suspected tampering was done.

To perform the assessment, the prototype will read the PDF using the PDFRW library and extract the file page and root objects. It will then extract the hash values stored in the root object associated with the following keys: 'hashroot' and 'hashinfo'. Following this, it will extract all the groups of hashes for each file page object in the PDF document from the relevant keys. The keys that will be extracted are: 'hashobject', 'hashroot', and 'hashleafs'.

After extracting the hash values from the root object and the file page objects, the keys will be removed from the respective objects before calculating the hash value for the relevant object. The calculated hash values will be compared to the previously extracted hash values to determine if the file has been tampered with.

A relevant message indicating the presence or absence of tampering will be displayed. Should tampering be detected in the PDF document, specifically within the root object, or the metadata of the PDF document, a message indicating this will be displayed. If tampering has been detected, the page number where the changes were made and in which group of the byte stream the changes will be displayed. Figure 5 shows the output of the assessment for forgery.

```
Would you like to protect a PDF? (y/n): n
Would you like to assess a PDF for tampering? (y/n): y
Enter the path to the PDF you would like to assess: ./Test_PDFs/Demo_hash.pdf
Assessing: ./Test_PDFs/Demo_hash.pdf
Hashes are equal, no tampering detected
```

**Fig. 5. Output when assessing for forgery**

## 4  Experimentation and Results

This section focuses on evaluating the prototype's effectiveness, which is crucial for determining its practicality and validity. It outlines the structure, starting with an explanation of changes made to PDFs for testing purposes, covering alterations to text, images, and metadata.

### 4.1  Results

The tests include scenarios such as text addition, alteration, and removal, image insertion, and manipulation, as well as metadata changes [14]. Each type of alteration is tested individually and then combined in a comprehensive test file. This thorough evaluation aims to assess the prototype's ability to detect various alterations accurately.

Tables 1 to 4 below tabulate the results of all the tests conducted using the prototype. All PDF documents used in the test cases described below were first protected by the prototype, and then the alterations were made. The only exception to this protection is the first test, which explicitly tests for the absence of the hashes. The PDF documents were created using Microsoft Word and Adobe Acrobat. PDF names with an asterisk (*) at the beginning denote that the PDF was created with Microsoft Word; those without were created with Adobe Acrobat. All changes to the text, images, and metadata within a PDF were done using Adobe Acrobat. The file page objects will change depending on the changes made to the text and images within a PDF document.

Table 1. Experimentation Results: Not Protected

| Test Name | PDF Name | Test Description |
|---|---|---|
| No Hashes | NoHash hash.pdf | This test assesses the ability of the prototype to detect the absence of hashes in a PDF. |
| Result | Assessing: ./Test_PDFs/NoHash.pdf<br>There was an error assessing the PDF:  No hash values found in PDF<br>Process Completed | |

Table 1 shows that the prototype can detect when a PDF does not have the hashes stored within the file. It also indicates that it cannot assess such a file because no hashes are stored.

Table 2. Experimentation Results: Text Alteration

| Test Name | PDF Name | Test Description |
|---|---|---|
| Single addition | * TextSA hash.pdf | This test sees the addition of a singular line of text on the 2nd page of the PDF document. |
| Result | Assessing: ./Test_PDFs/TextSA_hash.pdf<br>Hashes are not equal, alterations detected:<br><br>Root Hashes are not equal, root object has been altered<br>Info Hashes are not equal, metadata has changed<br>Object Hashes are not equal for page: 2<br>Root Hashes are not equal for page: 2<br>Changes detected in the 0 th 256 bytes of the content stream<br>Changes detected in the 1 th 256 bytes of the content stream | |
| Multiple addition | * TextMA hash.pdf | This test sees the addition of multiple singular lines of text on the 2nd page of the PDF document. |

Table 2 shows that the prototype can detect when the text in a PDF has been altered. Be this an addition, update, deletion, or a combination of all of the above. The messages displayed indicate which page or pages the text has been altered on and which parts of the content stream have been altered.

Table 3. Experimentation Results: Image Alteration

| Test Name | PDF Name | Test Description |
|---|---|---|
| Single addition | ImageSA hash.pdf | This test sees the addition of a singular image on the 2nd page of the PDF document. |
| Result | ```
Assessing: ./Test_PDFs/ImageSA_hash.pdf
Hashes are not equal, alterations detected:

Root Hashes are not equal, root object has been altered
Object Hashes are not equal for page: 2
Root Hashes are not equal for page: 2
Changes detected in the 0 th 256 bytes of the content stream
``` | |
| Multiple addition | ImageMA hash.pdf | This test sees the addition of multiple images on the 2nd and 3rd page of the PDF document. |
| Result | ```
Assessing: ./Test_PDFs/ImageMA_hash.pdf
Hashes are not equal, alterations detected:

Root Hashes are not equal, root object has been altered
Object Hashes are not equal for page: 2
Root Hashes are not equal for page: 2
Changes detected in the 0 th 256 bytes of the content stream
Object Hashes are not equal for page: 3
Root Hashes are not equal for page: 3
Changes detected in the 0 th 256 bytes of the content stream
``` | |

Table 3 shows results when the prototype successfully detects when an image or images in a PDF have been altered, whether it is an addition, update, deletion, or a combination of all of the above. The messages displayed indicate which page or pages the text has been altered on and which parts of the content stream have been altered.

Table 4. Experimentation Results: Metadata Alteration

| Test Name | PDF Name | Test Description |
|---|---|---|
| Single Update | * MetaSU hash.pdf | This test sees the addition of a singular metadata element in a PDF document. |
| Result | ```
Assessing: ./Test_PDFs/MetaSU_hash.pdf
Hashes are not equal, alterations detected:

Info Hashes are not equal, metadata has changed
``` | |



| Multiple Update | * MetaMU hash.pdf | This test sees the update of multiple metadata elements in the PDF document. |
|---|---|---|
| **Result** | Assessing: ./Test_PDFs/MetaMU_hash.pdf<br>Hashes are not equal, alterations detected:<br>Info Hashes are not equal, metadata has changed | |

Table 4 shows that the prototype can detect tampering with the metadata of a PDF document. The message displayed indicates that the metadata has been tampered with.

### 4.2   Evaluation

This section provides a critical evaluation of the prototype and its ability to detect tampering within a PDF document. First, we discuss the general ability of the prototype. Then we discuss the validity of the prototype's results and its limitations.

#### 4.2.1 General Analysis of the Prototype.

The prototype successfully detected changes in the three main categories we tested: changes to text, changes to images, and changes to metadata. These changes were all made using Adobe Acrobat. This means that we can only confirm that the prototype successfully detects changes when they are made using Adobe Acrobat. Different PDF editors produce PDF files with different layouts of underlying objects. However, the prototype should be able to detect the aforementioned changes regardless of the PDF editor, because the protected PDF documents are produced using the PDFRW library.

As seen in Section 3, in Tables 1 to 4, the output results describe on which page the change was detected or, in the case of metadata, that a change in the metadata object has been made. Further expansion on the indication of which page the change is on also indicates which parts of the content stream have been altered.

One thing to note is how the PDFRW library writes a PDF and the order of the objects in the PDF document differ from other PDF editors. For this reason, the file page objects, and the root object must be serialized before they can be used to produce a hash. In the current iteration of the prototype, specific sub-elements are selected and serialized for use in creating the hash for the object. It is worth noting that our proposed method works in instances when the PDF is protected (refer to Section 3.2). If it is not protected, then it would not be possible to detect where alterations were made.



**4.2.2 Evaluation of Prototype Validity.**
Since the objects must be serialized before they are used for the generation of the hash, any addition or alteration that does not affect the selected elements will not be detected by the prototype. Because all PDF editors save the PDF with a different underlying structure, we cannot serialize the entire object as-is because the order of these objects will impact the produced hash. This same issue with adding custom elements to a PDF can be applied to the PDF metadata, where one can add their custom metadata elements. The issue with not being able to consistently create a hash using the object is a complex task that requires extensive research into the most consistent way to convert it into a uniform input for the hash function.

When protecting the PDF document, a new file, a replica of the original that includes the hashes embedded into the file page objects, is created. This in itself is a form of tampering with the PDF document. There is no feasible way to insert the hashes into the original file without tampering. For the sake of later assessment by a document tampering specialist, making a copy of the original file was chosen as the route. This technique of protecting the PDF relies on the fact that only the protected PDF will be the one tampered with by a malicious individual.

We rely on the content stream for the majority of the detection of changes to the content of a particular PDF document. This will only detect a change to the physical text or the visible image to the user. If you were to change the font for a particular line or the entire document, then the prototype would not be able to detect this. This stems from the issue mentioned in the previous paragraph, where not all file page object elements can be used to generate the object's hash. While the content stream is part of the elements that make up a file page object, the current prototype uses its parent element when creating the hash for that particular file page object.

The prototype succeeds at detecting tampering with a PDF document. However, this is limited to text, images, and some types of alterations to metadata and file page objects. There are still many ways to modify the PDF document that the prototype would not detect, such as adding JavaScript to the PDF document.

## 5    Summary and Conclusion

The paper presented a technique to detect tampering or forgery in PDF documents using file page objects. The development of a prototype for this purpose is highlighted as a significant contribution. The developed prototype successfully utilized these file page objects to detect tampering within a PDF document. The file page objects were employed to generate hashes, which can be used to detect and analyze the presence of tampering within a PDF document.

Future research suggestions include generalizing the prototype to accommodate various PDF structures, investigating additional types of changes such as JavaScript additions and signature alterations, and considering the protection of scanned PDF



documents. These potential avenues aim to enhance the effectiveness and scope of tampering detection within PDF files.

# References


1. G. Saju and K. Sreenimol, "An effective method for detection and localization of tampering," International Journal of Information, vol. 8, pp. 152–154, no. 2, 2019.
2. P. Johnston and E. Elyan, "A review of digital video tampering: From simple editing to full synthesis," Digital Investigation, vol. 29, pp. 67–81, 2019.
3. T. Bienz, R. Cohn, and C. Adobe Systems: Mountain View, Portable document format reference manual. Citeseer, 1993.
4. N. A. A. Usop, S. I. Hisham, and J. M. Zain, "An implementing of zigzag pattern in numbering watermarking bits for high detection accuracy of tampers in document," Indian Journal of Computer Science and Engineering (IJCSE), vol. 13, pp. 1733-1751, 2022.
5. U. Khadam, M. M. Iqbal, M. A. Habib, and K. Han, "A watermarking technique based on file page objects for pdf," in 2019 IEEE Pacific Rim Conference on Communications, Computers and Signal Processing (PACRIM), pp. 1–5, IEEE, 2019.
6. I. B. Senkyire and Q.-A. Kester, "A cryptographic tamper detection approach for storage and preservation of forensic digital data based on sha-384 hash function," in 2021 International Conference on Computing, Computational Modelling and Applications (ICCMA), pp. 159–164, IEEE, 2021.
7. D.S. Popescu, "Hiding malicious content in pdf documents," Journal of Mobile, Embedded and Distributed Systems, 3 (3) (2011), pp. 120-127, 2012.
8. V. Mladenov, C. Mainka, K. Meyer zu Selhausen, M. Grothe, and J. Schwenk, "1 trillion-dollar refund: How to spoof pdf signatures," in Proceedings of the 2019 ACM SIGSAC Conference on Computer and Communications Security, pp. 1–14, 2019.
9. P. Dikanev and Y. Vybornova, "Method for protection of pdf documents against counterfeiting using semi-fragile watermarking," in 2021 International Conference on Information Technology and Nanotechnology (ITNT), pp. 1–4, IEEE, 2021.
10. C. Mainka, V. Mladenov, and S. Rohlmann, "Shadow attacks: Hiding and replacing content in signed pdfs.," in 28th Annual Network and Distributed System Security Symposium, NDSS 2021, The Internet Society, 2021.
11. C. Gao, X. Wan, C. Guo, B. Wu. "Blockchain-based PDF File Copyright Protection and Tracing." Springer Handbook of peer-to-peer networking, 2023.
12. N. Josue, F. Tchakounte, PM. Buhendwa, M. Atemkeng. "Fals-Ism: A Graph Isomorphism Framework for Multi-Level Detection of Falsified PDF Documents." Journal of Computer Science 19.5: pp. 667-676, 2023.
13. Z. Jiang, H. Wang, SY. Han. "A robust PDF watermarking scheme with versatility and compatibility." Multimedia Tools and Applications, pp. 1-27, 2024.
14. AH. Kamakshi, and S. Grandhi. "Text classification from PDF documents." International Research Journal of Modernization in Engineering Technology and Science 3: pp. 58-63, 2021.
15. Bitar, Ahmad W., et al. "Blind digital watermarking in PDF documents using Spread Transform Dither Modulation." Multimedia Tools and Applications 76: pp. 143-161, 2017.




16. A. Gholipour, & S, Mirzakuchaki. "High-speed implementation of the Keccak hash function on FPGA". International Journal of Advanced Computer Science, 2(8), pp: 303-307, 2012